\documentstyle[12pt]{article}
\def\lromn#1{\uppercase\expandafter{\romannumeral#1}}

\begin{document}

\begin{flushright}
TU/96/499\\
\end{flushright}

\vspace{12pt}

\begin{center}
\begin{Large}

\renewcommand{\thefootnote}{\fnsymbol{footnote}}
\bf{
Decay Rate of Coherent Field Oscillation 
}
\footnote[1]
{To appear in the Proceedings of the Symposium on 
"{\em Frontiers in Quantum Field Theory}",
(World Scientific, Singapore, 1996).
}

\end{Large}

\vspace{36pt}

\begin{large}
M. Yoshimura \\

Department of Physics, Tohoku University\\
Sendai 980-77 Japan\\
\end{large}

\vspace{54pt}

{\bf ABSTRACT}
\end{center}

\vspace{0.5cm} 
In recent studies it has become increasingly clear that presence of infinitely
many instability bands of the parametric resonance plays crucial roles
in the phenomenon of particle production under periodic classical 
field oscillation. We extend previous works to a general class of models 
including both the Yukawa and the
quartic type of couplings of the classical field to quantum bose fields. 
Decay rate from the $n-$th band is derived in the small amplitude limit
using the functional Schr$\stackrel{..}{{\rm o}}$dinger picture.
It is then shown that this analytic result of the decay rate
can also be derived as the zero momentum limit of a physical process,
$n$ particles that comprise the classical homogeneous field
decaying simultaneously into 2 bose particles.
The latter approach uses ordinary perturbation theory, hence the former result
is a novel resummation of many perturbative amplitudes, which usually becomes
complicated for a large $n$ order.

\newpage

{\bf 1 Introduction}

\vspace{0.5cm} 
\hspace*{0.5cm} 
Particle production under periodic perturbation is a physical process
that governs many problems in different areas of physics.
In cosmology in particular, it is a dominant process of entropy generation
after inflation, since inflation dilutes away essentially everything in
the observable part of our universe and leaves behind the field (inflaton)
oscillation \cite{linde et al 94}. 
As another example one may cite presence of flat field directions
in supersymmetirc models. It may either give good effects 
such as the Affleck-Dine scenario of baryogenesis \cite{affleck-dine}, 
or bad effects such as
the Polonyi or the modular problem \cite{polonyi} 
that may potentially destroy the successful result of nucleosynthesis.
In all these problems effects of the parametric resonance are crucial
if the initial oscillation amplitude is large enough.
Thus a deep understanding of the basic process is inevitable in any of these
applications.

In our previous study \cite{mine95-1}, \cite{fkyy95-1}
we proposed to formulate the problem of particle
production by taking a short time average of the quantum density matrix
over a few oscillation periods. In our view this time average is a substitute
for a more physical means of the 
coarse graining when created particles interact among themselves
or with some other particles. 
A large fluctuation of the quantum system coupled to
the classical periodic oscillation is essential to this viewpoint,
and the quantum system
exhibits a nearly classical behavior even before taking the time average.

In this work we bridge a gap by offering in the language of ordinary 
perturbation of interacting quantum field theory a simple understanding of the
decay formula previously derived.
We do this by generalizing the coupling of oscillating field to
quantum bose fields beyond that previously analyzed.
The model Lagrangian for the oscillator coupling is thus taken as
\begin{eqnarray}
{\cal L}_{{\rm int}}  = -\,\frac{1}{2}\,g_{4}^{2}\, \xi ^{2}\varphi ^{2} +
\frac{1}{2}\, g_{Y}m_{\xi }\,\xi \varphi ^{2} \,,
\end{eqnarray}
with $\xi $ the classical oscillating field that takes a simple sinusoidal
form,
\begin{eqnarray}
\xi (t)  = \xi _{0}\,\sin (m_{\xi }t) \,.
\end{eqnarray}
$m_{\xi }$ is the mass of the $\xi $ field, and 
$\varphi $ is a generic bose field treated here as a quantum field.
In the rest of discussion it is not important to assume a particular
relation between the two dimensionless couplings, the quartic coupling
$g_{4}$ and the Yukawa coupling $g_{Y}$,
but in order to organize results systematically we assume these to be
of the same order of magnitude; $g_{4}/g_{Y} = O[1]$.

The quantum state coupled to the external $\xi $ oscillation evolves
according to the functional Schr$\stackrel{..}{{\rm o}}$dinger equation,
which is much simplified to an independent set of equations in 
Fourier modes due to the 
translational invariance of the background  oscillation.
Thus the state vector in each $\vec{k}$ mode obeys the 
Schr$\stackrel{..}{{\rm o}}$dinger equation with variable frequency,
\begin{eqnarray}
\omega _{k}^{2}(t) = \vec{k}^{2} + g_{4}^{2}\,\xi ^{2}(t) -
g_{Y}m_{\xi }\,\xi (t) \,.
\end{eqnarray}
Precise relation between the quantum wave function $\psi_{k} (q_{k}\,, t)$
and the classcial oscillator equation has been established \cite{mine95-1}. 
It takes
the most definitive form if one starts with an initial quantum state of
the ground state of some reference frequency $\omega $;
\( \:
\psi _{k}  (q_{k}\,, t) =  (\omega /\pi )^{1/4}\,e^{-\,\frac{i}{2}\,
\omega t}\,\exp [-\,\frac{1}{2}\, \omega q_{k}^{2}] \,,
\: \)
for $t \approx 0$.
At any finite time $t$
\begin{eqnarray}
\psi _{k}  (q_{k}\,, t) =
\frac{1}{\sqrt{|u_{k}(t)|}}\,\exp [\,-\,\frac{\pi }{2|u_{k}(t)|^{2}}\,q_{k}^{2}
+ \frac{i}{4}\frac{d}{dt}\,\ln |u_{k}(t)|^{2}\cdot q_{k}^{2}\,] \,,
\end{eqnarray}
where the time dependent complex function $u_{k}(t)$ obeys 
the  classical oscillator equation with the definite initial condition,
\begin{eqnarray}
\frac{d^{2}u _{k}}{dt^{2}} + \omega_{k} ^{2}(t)\,u _{k} = 0 \,,
\hspace{0.5cm} 
u_{k} (0) = (\omega /\pi )^{-1/2}\,, \hspace{0.5cm} 
\dot{u}_{k}(0) = i\omega \,u_{k} (0) \,.
\end{eqnarray}

The meaning of the classical oscillator amplitude $u_{k}(t)$ is thus
unambiguous: 
its modulus $|u_{k}|$ governs the Gaussian width of the wave function 
and at the same time its logarithmic derivative gives the fluctuation of
the state in terms of the phase factor. 
This can be seen most clearly in the Fock
space base of harmonic oscillator of the reference frequency: the diagonal
density matrix element derived from this wave function has a smallest
fluctuation and is related to the average particle number 
$\langle  N_{\omega } \rangle$ according to 
\( \:
\rho _{2n \,, 2n} \:\rightarrow  \: e^{-\,n/\langle N_{\omega } \rangle}/
\sqrt{\,\pi n\,\langle N_{\omega } \rangle\,} 
\: \)
\cite{mine95-1}, 
while the off-diagonal density matrix elements contain wildly varying
signs of $\pm $, giving zero after taking the time 
average over a few oscillation periods.
This phenomenon occurs in infinitely many band regions 
of the parameter space $(k\,, \xi _{0})$ within which  a generic classical
$u_{k}$ exponentially grows; 
\( \:
u_{k}(t) \:\rightarrow  \: e^{\lambda m_{\xi }t/2} \times 
\: \)
(periodic function) with $\lambda > 0$.
The exponential growth implies rapid excitation of high harmonic
oscillator levels. With the short time average it can be interpreted
that particle production takes place with
\( \:
\langle N_{\omega } \rangle \:\propto  \: e^{\lambda m_{\xi }t} \,.
\: \)

After coarse graining of the time average one can discuss the decay law
of the initially prepared ground state: it follows the exponential form,
\begin{eqnarray}
\rho _{00} \approx e^{-\,\Gamma \,Vt} \,, \hspace{0.5cm} 
\Gamma = \sum_{n = 1}^{\infty }\,\Gamma _{n} \,, \hspace{0.5cm} 
\Gamma _{n} = \frac{m_{\xi }}{2V}\,
\sum_{\vec{k}\, \in \,n-{\rm band}}\,\lambda _{\vec{k}}\,.
\end{eqnarray}
$\Gamma $ is the total decay rate per unit volume and per unit time.
Computation and interpretation of the decay rate $\Gamma _{n}$ of 
the $n-$th band is our main task in the rest of discussion.

\vspace{0.5cm} 
{\bf 2 Small Amplitude Analysis Revisited}

\vspace{0.5cm} 
\hspace*{0.5cm} 
It is customary to recast the classcial equation into a dimensionless form,
\begin{eqnarray}
&&
\frac{d^{2}u}{dz^{2}} + [\,h - 2\theta _{4}\cos (4z) - 
2\theta _{Y}\sin (2z)\,]u = 0 \,, \\
&& \hspace*{-1.5cm}
z = m_{\xi }t/2 \,, \hspace{0.5cm} 
h = 4\frac{\vec{k}^{2} + m^{2}}{m_{\xi }^{2}} + 2
\frac{g_{4}^{2}\xi _{0}^{2}}{m_{\xi }^{2}} \,, \hspace{0.5cm} 
\theta _{4} = \frac{g_{4}^{2}\xi _{0}^{2}}{m_{\xi }^{2}} \,, \hspace{0.5cm} 
\theta _{Y} = 2\frac{g_{Y}\xi _{0}}{m_{\xi }} \,,
\end{eqnarray}
with $m$ the mass of the bose quantum field $\varphi $.
Following the general theorem for solutions of differential equation
with periodic coefficients, one expands the solution in the form 
\cite{coddington},
\begin{eqnarray}
u(z)  = \sum_{k = -\infty }^{\infty }\,c_{k}\,e^{(\lambda + in + 2ik)z} \,,
\end{eqnarray}
with 
\( \:
n = 1 \,, 2\,, 3\,, \cdots .
\: \)
The condition of the existence of non-trivial solution is then 
equivalent to the
condition of non-vanishing matrix determinant of infinite dimensions,
with the diagonal entry of $\gamma _{k} = h + (\lambda + in + 2ik)^{2} $,
and the next off-diagonal entry of $\pm i\theta _{Y}$ and the next-to-next
off-diagonal entry of $-\,\theta _{4}$ and all others $= 0$.

In the small amplitude limit of $|\xi _{0}| \ll 1$ the $n-$th instability band
starts at $h = n^{2}$. The boundary curve $h = h^{(n)}(\xi _{0}/m_{\xi })$
dividing the stable and the unstable bands
corresponds to the eigen-modes of the form, $\cos (nz)$ and
$\sin (nz)$, in the $\xi _{0} \:\rightarrow  \: 0$ limit.
With the assumption that the coefficients, $c_{0}$ and $c_{-n}$, 
are dominant and 
the rest of $c_{k}$, $\lambda $, and $h - n^{2}$ are all small,
one may simplify the structure of this matrix such that
\( \:
\gamma _{k} = \gamma _{-n-k} = -\,4k(n + k) \; (k \neq 0 \,, -n)
\: \)
and 
\( \:
\gamma _{0} = h - n^{2} + 2in\lambda \,, \hspace{0.5cm} 
\gamma _{-n} = h - n^{2} - 2in\lambda \,.
\: \)
Dividing the matrix into a few parts, one first solves the top 
($k \geq 1$) and the down ($k \leq -n-1$) infinite dimensional parts 
in favor of $c_{1} \,, c_{2}$ and $c_{-n-1} \,, c_{-n-2}$,
\begin{eqnarray}
&& \hspace*{-1cm}
c_{2}  = D^{-1}_{21}(\epsilon _{1}c_{0} + \epsilon _{2}c_{-1}) +
D^{-1}_{22}\epsilon _{2}c_{0} \,, \hspace{0.5cm} 
c_{1}  = D^{-1}_{11}(\epsilon _{1}c_{0} + \epsilon _{2}c_{-1}) +
D^{-1}_{12}\epsilon _{2}c_{0} \,, \nonumber \\
&& 
c_{-n-1} = D^{-1}_{11}(\epsilon _{1}^{*}c_{-n} + \epsilon _{2}^{*}c_{-n+1}) +
D^{-1 *}_{12}\epsilon _{2}^{*}c_{-n} \,, \nonumber \\
&&
c_{-n-2} = D^{-1 *}_{21}(\epsilon _{1}^{*}c_{-n} + \epsilon _{2}^{*}c_{-n+1})
+ D^{-1}_{22}\epsilon _{2}^{*}c_{-n} \,,
\end{eqnarray}
with $\epsilon _{1} = -i\theta _{Y}$ and $\epsilon _{2} = \theta _{4}$
and the matrix inverse $D^{-1}$ defined as a limit of big matrix in
the left-upper and the right-down (identical) corners.
One next solves the central block ($ -1 \geq  k \geq -n+1$) for
$c_{-1} \,, c_{-2} \,, c_{-n+2}\,,  c_{-n+1}$ 
in terms of $c_{0} \,, c_{-n}$.
Here one needs to invert, ignoring subleading terms,
\begin{eqnarray}
&& \hspace*{-2cm}
\left( 
\begin{array}{ccccccc}
\gamma _{-1} & -\,\epsilon _{1} & -\epsilon _{2}& 0 & \cdots & \cdots & 0  \\
-\,\epsilon _{1}^{*} & \gamma _{-2} & -\,\epsilon _{1} & -\epsilon _{2}& 0 & 
\cdots & 0 \\
-\,\epsilon _{2}^{*} & \cdots & \cdots & \cdots &\cdots & \cdots & 0 \\
0 & \cdots & \cdots & \cdots &\cdots & \cdots & 0 \\
0 & \cdots & \cdots & \cdots &\cdots & \cdots & -\,\epsilon _{2} \\
0 & \cdots  & 0 & -\,\epsilon _{2}^{*}& -\,\epsilon _{1} ^{*}
& \gamma _{-n+2} & -\,\epsilon _{1} \\
0 & \cdots & \cdots  & 0 & -\,\epsilon _{2}^{*} & -\,\epsilon _{1}^{*} 
& \gamma _{-n+1}
\end{array}
\right) 
\,
\left( 
\begin{array}{c}
c_{-1} \\ c_{-2} \\ \cdot \\  \cdot \\ \cdot \\
c_{-n+2} \\ c_{-n+1}
\end{array}
\right) = 
\left( 
\begin{array}{c}
\epsilon _{1}^{*}c_{0} \\ \epsilon _{2}^{*}c_{0}  \\
0 \\ \cdot \\ 0 \\ \epsilon _{2}c_{-n} \\ \epsilon _{1}c_{-n}
\end{array}
\right) \,. \nonumber \\ \label{central block eq} 
\end{eqnarray}

One finally solves the determinental condition for the equation 
$c_{0} \,, c_{-n}$, which reads to the leading order as
\begin{eqnarray}
&& \hspace*{0.5cm} 
{\rm det}\,
\left( 
\begin{array}{cc}
h - n^{2} + 2in\lambda - A 
&  -\,C \\
-\,C^{*} & h - n^{2} - 2in\lambda - A
\end{array}
\right) = 0 \,, \\
&& \hspace*{-1.5cm}
A = |\epsilon _{1}|^{2}(D^{-1}_{11} + E^{-1}_{11}) \,, \hspace{0.3cm} 
C = \epsilon _{1}(\epsilon _{1}E^{-1}_{1\,, n-1} + \epsilon _{2}E^{-1}_{1
\,, n-2}) + \epsilon _{2}(\epsilon _{1}E^{-1}_{2\,, n-1} + \epsilon _{2}
E^{-1}_{2\,, n-2}) \,, \nonumber \\
\end{eqnarray}
where $E$ is the matrix in the left hand side of eq.\ref{central block eq}.

We regard $\epsilon _{1} = O[\epsilon ]$ and 
$\epsilon _{2} = O[\epsilon ^{2}]$ since $g_{4}/g_{Y} = O[1]$, 
and work out matirix elements to
leading orders of $\epsilon $.
The net result may  be summarized as the formula for the growth rate
$\lambda $,
\begin{eqnarray}
&& \hspace*{-1cm}
\lambda _{n} = \frac{1}{2n}\,\sqrt{\,\Delta_{n} ^{2} - 
(h - n^{2} - \frac{\theta _{Y}^{2}}{2(n^{2} - 1)})^{2}\,}
\,, \hspace{0.5cm} 
\Delta _{n} = \frac{|{\cal C}_{n}|}{2^{2(n-1)}\,[(n-1)!]^{2}}\,, \\
&&
{\cal C}_{n} = {\rm det}\,
\left( 
\begin{array}{cccccc}
-\,i\theta _{Y} & \theta _{4} & 0 & \cdots & \cdots & 0  \\
-\,\gamma _{-1} & -\,i\theta _{Y} & \theta _{4} & 0 & 
\cdots &  0 \\
0 & \cdots & \cdots &\cdots & \cdots & 0 \\
0 & \cdots & \cdots &\cdots & \cdots & 0 \\
0 & \cdots  & 0 & -\,\gamma _{-n+2} & -\,i\theta _{Y} & \theta _{4}
\\
0 & \cdots & \cdots  & 0 & -\,\gamma _{-n+1} & -\,i\theta _{Y}
\end{array}
\right) \,, \label{y4matrix}
\end{eqnarray}
with $\gamma _{-k} = 4k(n-k)$.
This result generalizes the previous one in ref \cite{mine95-1} 
in which $\theta _{4} = 0$
was assumed, hence ${\cal C}_{n} = (-\,i\theta _{Y})^{n}$.
The decay rate of the $n-$th band is computed by summing modes within the
band in the narrow width approximation,
\begin{eqnarray}
\Gamma _{n} = \frac{m_{\xi }^{4}}{256\pi }\,\sqrt{\,1 - \frac{4m^{2}}
{n^{2}m_{\xi }^{2}}\,}\,\Delta_{n} ^{2} \,. \label{rate formula} 
\end{eqnarray}

The case of $n =1$ is an exception to this general formula that
must be treated separately, resulting in
\begin{eqnarray}
\lambda _{1}(k\,,\xi _{0})  = \frac{1}{2}\, \sqrt{\, \theta _{Y}^{2} - 
(h - 1 + \frac{\theta _{Y}^{2}}{8})^{2}\,} \,, \hspace{0.5cm} 
\Gamma _{1} = 
\frac{g_{Y}^{2}m_{\xi }^{2}\xi_{0} ^{2}}{64\pi }\,\sqrt{\,1 - \frac{4m^{2}}
{m_{\xi }^{2}}\,} \,.
\end{eqnarray}

\vspace{0.5cm} 
{\bf 3 Physical Interpretation in terms of Familiar Perturbation Theory}

\vspace{0.5cm} 
\hspace*{0.5cm} 
We shall now offer interpretation of the decay rate $\Gamma _{n}$
of $n-$th band in terms of the ordinary perturbation theory.
Ordinary perturbation uses the particle picture and for that purpose
it is useful to recast the rate formula in terms of reaction rates of
indivisual particles by
dividing some powers of the particle number density $n_{\xi } =
\frac{1}{2}\, m_{\xi }\xi _{0}^{2}$.
Let us first note the dependence of the rate $\Gamma _{n}$ on the 
oscillation amplitude, $ \propto \xi _{0}^{2n}$, and
the momentum of final $\varphi $ particles in the zero momentum limit
of $E_{\xi } \rightarrow m_{\xi }$ for the process,
\( \:
n\,\xi \:\rightarrow  \: \varphi \,\varphi \,;
\: \)
\( \:
p_{\varphi } = \frac{nm_{\xi  }}{2}\,\sqrt{1 - \frac{4m^{2}}
{n^{2}m_{\xi }^{2}}}
\: \).
This factor appears in the rate formula, eq.\ref{rate formula}.

The simplest case is the decay rate of the first band $\Gamma _{1}$,
the rate per unit volume and per unit time. Since the unit volume
contains $n_{\xi }$ particles, the one particle rate having
the dimension of the inverse time is \cite{mine95-1}
\begin{eqnarray}
\frac{\Gamma _{1}}{n_{\xi }} = 
\frac{g_{Y}^{2}m_{\xi }}{32\pi }\,\sqrt{\,1 - \frac{4m^{2}}
{m_{\xi }^{2}}\,} \,.
\end{eqnarray}
This exactly coincides with the decay rate of one $\xi $ particle computed 
in the conventional way using the Yukawa coupling to $\varphi $,
\( \:
\frac{1}{2}\, g_{Y}m_{\xi }\,\xi \varphi ^{2}
\: \).

The next thing to be checked is the decay rate from the 2nd instability
band $\Gamma _{2} \:\propto  \: \xi _{0}^{4}$.
This time one divides the quantity $\Gamma _{2}$ 
by $n_{\xi }^{2}$ since two $\xi $ particles are involved,
\begin{eqnarray}
\frac{\Gamma _{2}}{n_{\xi }^{2}} = \frac{(g_{4}^{2} - g_{Y}^{2})^{2}}{16\pi }
\,\frac{1}{m_{\xi }^{2}}\,\sqrt{1 - \frac{m^{2}}{m_{\xi }^{2}}} \,.
\end{eqnarray}
In ordinary perturbation theory the amplitude for the 2-body process,
$\xi \xi \:\rightarrow  \: \varphi \varphi $, consists of 3 distinct
Feynman amplitudes, the contact term with a single quartic coupling, 
the t-channel and the u-channel exchange diagrams with two vertices of
the Yukawa coupling, adding to
\begin{eqnarray}
-\,2ig_{4}^{2} - ig_{Y}^{2}m_{\xi }^{2}(\frac{1}{t - m^{2}} +
\frac{1}{u - m^{2}}) \,.
\end{eqnarray}
In the zero momentum limit of $E_{\xi } \:\rightarrow  \: m_{\xi }$,
\( \:
t - m^{2} = u - m^{2} \:\rightarrow  \: -\,m_{\xi }^{2},
\: \)
giving the total amplitude $2i(g_{Y}^{2} - g_{4}^{2})$.
Working out the phase space factor, one finds out that the invariant rate
$v_{{\rm rel}}\,\sigma $
given by flux $\times $ cross section is identical to
$\Gamma _{2}/n_{\xi }^{2}$ given above.

To proceed to the general $n-$th order case, 
one notes that the propagator in the zero momentum limit is given by
\begin{equation}
\frac{i}{(\sum_{i = 1}^{k}\,p_{i} - q_{1})^{2} - m^{2}} \:\rightarrow  \:
\frac{-\,i}{k(n - k)\,m_{\xi }^{2}} \,,
\end{equation}
where $p_{i}$ is an initial $\xi $ momentum and $q_{1}$ is one of
the final $\varphi $ momenta.
Hence the $n-$th order invariant amplitude containing 
the Yukawa couplings alone is
\begin{eqnarray}
\frac{i\,(g_{Y}m_{\xi })^{n}}{m_{\xi }^{2(n-1)}}\,\prod_{k=1}^{n-1}\,
\frac{1}{k(n-k)} = \frac{ig_{Y}^{n}}{m_{\xi }^{n-2}}\,\frac{1}{[(n-1)!]^{2}}
\,.
\end{eqnarray}
It is not difficult to check that this leads to the invariant rate 
precisely equal to
the corresponding decay rate of the $n-$th band, $\Gamma _{n}/n_{\xi }^{n}$.
What is left to be shown is then the relative weight of the Yukawa and the
quartic contribution. Subdiagrams of $\xi \xi \rightarrow 
\varphi \varphi $ for the whole $n\xi \rightarrow \varphi \varphi $
process contribute with a factor,
$\frac{ig_{Y}^{2}}{k(n-k)}$ for the Yukawa coupling case,
considering the propagator above, and with $-ig_{4}^{2}$ for the
contact quartic coupling case. 
The ratio of these two terms is exactly equal to the
ratio in
\begin{equation}
\hspace*{-0.5cm}
{\rm det} \,\left( \begin{array}{cc}
-2ig_{Y}\frac{\xi _{0}}{m_{\xi }} &  g_{4}^{2}\frac{\xi _{0}^{2}}{m_{\xi }^{2}}
\\
-\,4k(n-k) & -2ig_{Y}\frac{\xi _{0}}{m_{\xi }}
\end{array}
\right) =
-\,4k(n-k)\,
[\,\frac{g_{Y}^{2}}{k(n-k)} - g_{4}^{2}\,]\,(\frac{\xi _{0}}{m_{\xi }})^{2} \,,
\end{equation}
that appears in the decomposition of the submatrix of ${\cal C}_{n}$, 
eq. \ref{y4matrix} for the formula $\Gamma _{n}$.
This proves our assertion that the decay rate of the $n-$th band
$\Gamma _{n}/n_{\xi }^{n}$ is equal to the zero momentum limit of the
invariant rate computed in the ordinary perturbation theory.

Needless to say, this interpretation is valid only in the small amplitude
limit. In applications to realistic problems analytic formula in the
large amplitude regime is indispensable.
In cosmological application the parameter region along $h = 2\theta _{4}$,
or more precisely
\( \:
h - 2\theta _{4}\, (= 4\frac{\vec{k}^{2} + m^{2}}{m_{\xi }^{2}}) \ll 
\theta _{4}
\: \)
with the large amplitude of $\theta _{4} \gg  1$ 
is most important, whose case has been worked
out in ref \cite{fkyy95-1}.
In the present note we clarified the physical meaning of the $n-$th band decay
formula when it is taken to the $\xi _{0} \rightarrow 0$ limit.
In view of our result here 
the large amplitude formula in ref \cite{fkyy95-1} may be considered
non-perturbative effect which is directly joined to the present 
perturbative result.

\newpage

\end{document}